\documentclass[final,5p,times]{elsarticle}
\journal{Physica A}

\usepackage{graphics,graphicx,epsfig,color,latexsym,float,amsmath,amssymb,bm}
\usepackage[colorlinks,citecolor=blue,linkcolor=blue,urlcolor=blue]{hyperref}
\usepackage[version=4]{mhchem}

\usepackage{tikz}
\usepackage{tikzscale}
\usetikzlibrary{positioning, calc}
\usepackage{wrapfig}

\usepackage{dcolumn}
\usepackage[caption=false]{subfig}
\usepackage{csvsimple}

\usepackage{newfloat}
\usepackage{listings}

\DeclareFloatingEnvironment[
	fileext=lol,
	listname={List of Listings},
	name=LISTING,
	placement=htbp
]{lstfloat}

\lstdefinestyle{spice}{
	backgroundcolor=\color{white},   
	commentstyle=\color{green},
	keywordstyle=\color{magenta},
	stringstyle=\color{purple},
	basicstyle=\ttfamily\scriptsize,
	frame=single,
	breakatwhitespace=false,         
	breaklines=true,                 
	captionpos=b,                    
	keepspaces=true,                 
	numbers=left,                    
	showspaces=false,                
	showstringspaces=false,
	showtabs=false,                  
	tabsize=2
}
\begin{document}

\begin{frontmatter}

\title{A superstatistics approach to the modelling of memristor current--voltage responses}

\author[RU]{Roland Konlechner}
\author[UAE,US]{Anis Allagui}
\author[RU,UK]{Vladimir N. Antonov}
\author[RU]{Dmitry Yudin}
\affiliation[RU]{organization={Skolkovo Institute of Science and Technology},
            city={Moscow},
            postcode={121205}, 
            country={Russia}}
\affiliation[UAE]{organization={Department of Sustainable and Renewable Energy Engineering, University of Sharjah},
            city={Sharjah},
            postcode={27272}, 
            country={United Arab Emirates}}
\affiliation[US]{organization={Department of Mechanical and Materials Engineering, Florida International University},
            city={Miami},
            postcode={33174}, 
            state={FL},
            country={United States}}
\affiliation[UK]{organization={Royal Holloway, University of London},
            city={Egham, Surrey},
            postcode={TW20 0EX}, 
            country={United Kingdom}}

\begin{abstract}
Memristors are expected to form a major cornerstone in the upcoming renaissance of analog computing, owing to their very small spatial footprint and low power consumption. Due to the nature of their structure and operation, the response of a memristor is intrinsically tied  to local variabilities in the device. This characteristic is amplified by currently employed semiconductor fabrication processes, which introduce spatial inhomogeneities into the structural fabric that makes up the layers of  memristors.  In this work, we propose a novel $q$-deformed current--voltage model for memristors based on the  superstatistics framework, which allows the description of system-level responses while taking local variabilities into account. Applied on a Ag--Cu based synaptic memory cell, we  demonstrate that our model has a 4--14\% lower error than currently used models. Additionally, we show how the resulting $q$-parameter can be used to make statements about the internal makeup of the memristor, giving insights to spatial inhomogeneities and quality control.
\end{abstract}





\end{frontmatter}

\section{Introduction}
The memristor is a two-terminal device, first conceptualized by L.~Chua back in 1971~\cite{chua_memristor}; its deliberate fabrication however only began a few decades later~\cite{williams_memristor_found}. The distinguishing feature of the memristor is the ability to change its resistance in a non-volatile manner, dictated by an internal state. Due to this characteristic, memristors are believed to have far-reaching implications for the upcoming generations of computing systems. In particular, their nonvolatile nature, small footprint and stackability allow them to be used as a high-density RRAM~\cite{memristor_sram, memristor_4t2r_tcam,memristor_flip_flop}. Arranged in a crossbar array, memristors can store weights and perform matrix multiplications directly in hardware, which is highly relevant for machine learning applications~\cite{memristor_crossbar_logic, memristor_devices_for_computing, memristor_cnn, nn_crossbar_mapping, memristor_gradient_update_algorithm}. Moreover, thanks to their similarity to biological neurons, memristors are suitable to simulate synaptic memory cells in neuromorphic computing applications~\cite{yakipcic_neuromorphic_application, jo-nanoscale-memristor, memristor_neuromorphic_computing, ilyas_ag_siox_ag_tiox_mem_device, memristors_energy_efficient_computing_paradigms}. Computer designs incorporating memristors can circumvent the von Neumann bottleneck, solve the ever-more apparent limitations of current CMOS technology and have the prospect of reducing the power consumption for computing operations by orders of magnitude~\cite{memristor_gradient_update_algorithm,memristors_energy_efficient_computing_paradigms,von_neumann_bottleneck}.

Many attempts to model memristive devices available in the literature nowadays treat the memristor as a black box, applying purely electrical reasoning as a means of characterization~\cite{williams_memristor_found, joglekar_window_function, biolek_window_function, prodromakis_window_function, spice_modelling_memristors, memristive_switching_mechanism, memristor_team, memristor_chang, generalized_memristor_model, abu_bakar_model, updated_generalized_model, memristor_charge_transfer_model}. However, the fundamental mechanism of memristive switching is inherently tied to the internal makeup of the device. The main switching property, as well as other characteristics such as device degradation, are all direct consequences of the dynamic evolution of the internal memristor structure~\cite{comprehensive_memristor_review, controllable_growth_nanofilaments, atomic_structure_nanofilaments, electrochemical_dynamics_of_metallic_inclusions, understanding_memristive_switching, jo_programmable_resistance_switching, memristor_physics_part_1, memristor_filament_degradation}. This natural structural dependency, together with common methods of memristor fabrication widely used at the moment~\cite{optical_properties_co_sputtering, cantor-microstructore-si-ag-nanocomposites}, makes those devices intrinsically very stochastic. Since memristors are considered to be at the heart of many prospected technological innovations, it is important to describe those devices adequately and in a computationally economical manner, taking their statistical variability into account. In particular,  knowing that one might identify up to nine different operating mechanisms for memristive devices~\cite{comprehensive_memristor_review}, the need to develop an accurate memristor model is becoming paramount  for any further development based on this technology.

In this work, the superstatistical approach of Beck and Cohen~\cite{superstatistics_2003} will be used to develop a mean-field model for the  current--voltage characteristic of synaptic memory cells. Superstatistics, which itself is derived from the Bayesian statistical analysis, can be seen as a generalization of the ubiquitous Boltzmann-Gibbs statistics, while being able to explain and give insights about complex dynamic systems away from equilibrium~\cite{superstatistics_2003, superstatistics_applications_2004, superstatistics_cohen_2007, superstatistics_wind_power, power_law_chemical_reaction, power_grid_superstatistics, deformed_bv_equations}. The use of the superstatistical approach in the context of this research is motivated by the fact that our fabricated Ag--Cu based synaptic memory cells, taken as a case study, should be viewed as a system which operates far from equilibrium, exhibiting multiple local response time constants due to microscopic inhomogeneities and irreversibilities. Due to these inhomogeneities, the overall response of the device can be interpreted as the superposition of several statistics of different scales. To the authors' best knowledge, superstatistical principles have not yet been employed as a means to model memristors, which constitutes the main contribution of this work to existing studies. 

The rest of the paper is organized as follows. In Section~\ref{sec:background}, we give a brief description of the internal structure, operating mechanism, and fabrication strategies used for filament-based memristors. We also review some of the widely used electrical models for memristor characterization and cover elements of the superstatistics approach by Beck and Cohen for modelling non-equilibrium systems~\cite{superstatistics_2003}. In Section~\ref{sec:methodology}, we describe our research methodology in terms of experimental procedure for data collection and numerical analysis used for data fitting. The development of the $q$-deformed current--voltage model, its validation and  discussion of its significance are given in Section~\ref{sec:rnd}. We conclude with our final remarks and highlight some future research possibilities in Section~\ref{sec:conclusion}.

\begin{figure}
		\centering
			\begin{tikzpicture}
		\node[draw=gray!50!blue, fill=gray!50!blue, thick, minimum height=0.6cm, minimum width=2cm](anode) at (0,0) {\color{white}{Anode: Al}};
		\node[draw=gray!60!red, fill=gray!60!red, thick, minimum height=0.6cm, minimum width=2cm, below = 0cm of anode](dopant) {\color{white}{Ag--Cu}};
		\node[draw=black!30, fill=black!30, thick, minimum height=0.6cm, minimum width=3cm, below = 0cm of dopant](carrier){\color{white}{intrinsic Si}};
		\node[draw=black!60, fill=black!60, thick, minimum height=0.6cm, minimum width=3cm, below= 0cm of carrier](Cathode: ){\color{white}{$p^+$ Si wafer}};
	\end{tikzpicture}
		\includegraphics[width=.625\linewidth]{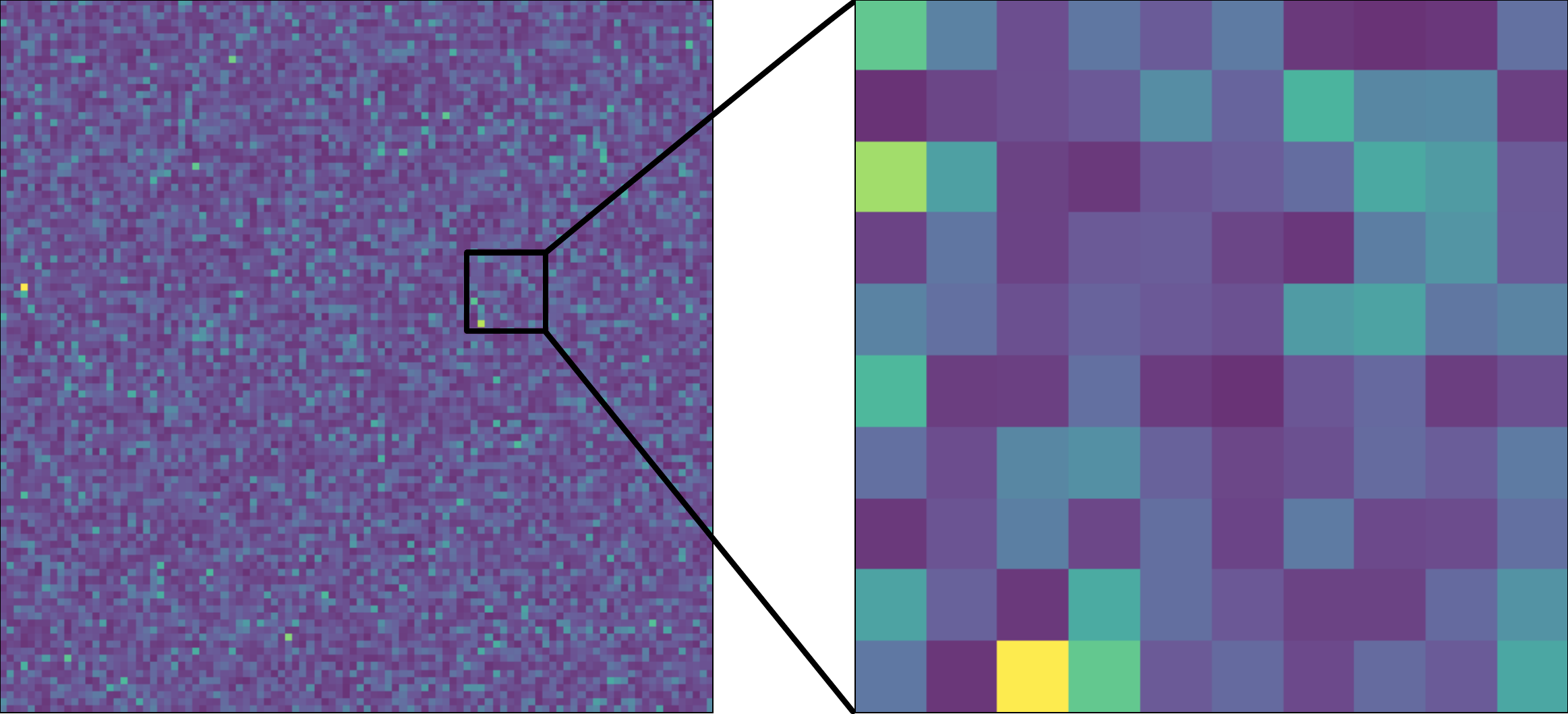}
		\caption{(left) Schematic of fabricated memristor device; (right) Illustration of a system with a fluctuating intensity parameter $\beta$, spatially divided into local cells. Drawn from $f(\beta)=\beta e^{-\beta/2}/4$, {\it i.e.}, $b=c=2$ in \eqref{eq:generalized_boltzmann_gamma_0}.}
		\label{fig:gamma_noise_closeup}
\end{figure}
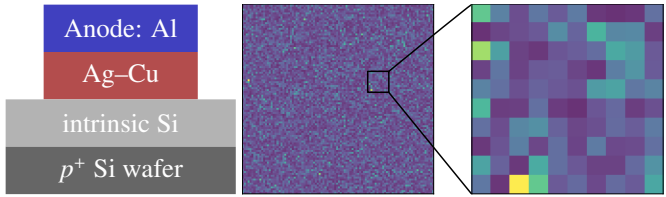

\section{Background}\label{sec:background}
\subsection{Filament-based memristor operation and fabrication}
At the device level, when a voltage is applied to Ag--Cu based memristors (Fig.~\ref{fig:gamma_noise_closeup}), the resulting electric field within the isolating layer causes the mobile metallic dopants to be ionized. These resulting ions are able to migrate within the carrier substrate, forming conducting channels or {\it filaments}. Once a filament touches both electrodes, the device switches from a high resistance state (HRS) to a low resistance state (LRS). Upon applying a voltage in the reverse direction, the filament ruptures, switching the state back from LRS to HRS. The forming of those conductive channels has been observed and reported in several studies~\cite{controllable_growth_nanofilaments, atomic_structure_nanofilaments, electrochemical_dynamics_of_metallic_inclusions, understanding_memristive_switching, jo_programmable_resistance_switching}. In the forming process, filaments tend to stay very thin and can rupture on their own, even without applying a reverse electric field~\cite{silver_nanofilament, surface_diffusion_limited_lifetime}. Over many actuation cycles, filament-based memristive devices tend to degrade, as the isolating layer between the electrodes forms parasitic conducting channels~\cite{memristor_physics_part_1,memristor_filament_degradation}. From an electrical standpoint, the current conduction mechanisms of filament-based memristive devices can be explained in various ways. Common explanations which are often found in the literature include Ohmic conduction, Simmons tunnelling, or Schottky emission~\cite{conduction_mechanism_survey}.

The fabrication of filament-based memristive devices utilizes common methods adapted from semiconductor manufacturing processes. Most frequently, introducing mobile dopants into the carrier substrate is achieved by a means of {\it co-sputtering} \cite{jo-nanoscale-memristor, ilyas_ag_siox_ag_tiox_mem_device}. This process does not produce an even distribution of dopants in the substrate, but forms nanoparticles. Nanoparticle formation is encouraged with an increasing amount of dopant in the substrate. Over time, the nanoparticles have the tendency to clump together and form fewer, larger droplets in the substrate. The spatial distribution of dopants can therefore be understood as a function of saturation and time, a process which is explained by the Rayleigh instability~\cite{optical_properties_co_sputtering, cantor-microstructore-si-ag-nanocomposites, silver_nanofilament}. Bright-field transmission electron microscopy images of this inhomogeneous spatial distribution --- due to clustering and surface segregation --- can be clearly seen for example in Figs.~1 and 4 in Ref.~\cite{cantor-microstructore-si-ag-nanocomposites} for the case of nanosized Ag particles embedded in an amorphous Si matrix prepared by the RF co-sputtering technique.

\subsection{Memristor modelling}
Many memristor models have been introduced to the literature over the years, which can be roughly divided into two approaches: physicochemical and electrical modelling. While physicochemical models try to simulate the dynamic evolution of the internal memristor structure~\cite{understanding_memristive_switching,surface_diffusion_limited_lifetime,kmc_simulation_of_hfo2,kmc_filament_thickening,kmc_simulation_electroforming_set_reset}, electrical models apply our current understanding of electrical conduction mechanisms to generate a macroscopic view of the memristor state change. Although physicochemical models are more accurate than electrical models, they are computationally very demanding, allowing only for single memristive devices to be studied. Electrical models on the other hand are less accurate, but also allow for constellations of many memristors to be simulated. This work will constitute a combination of the two approaches, incorporating a stochastic, macroscopic interpretation of physicochemical modelling into electrical models.

The first widely referenced system-level model of memristors was proposed by Strukov {\it et al.}~\cite{williams_memristor_found} which assumes two regions, doped and undoped, with respective Ohmic resistances. This model has been extended by window functions~\cite{joglekar_window_function, biolek_window_function, prodromakis_window_function} to capture nonlinear ion drift phenomena which are often found in observations. Later, Yang {\it et al.} introduced an updated $i$--$v$ equation, which was specified by five parameters to be fitted to experimental data --- the resulting model explicitly accounts for the electron tunneling and rectifying effect. The tunnelling mechanism explains electric current through metal-insulator-metal (MIM) junctions with a thin dielectric film~\cite{memristive_switching_mechanism}, and has since been widely accepted as one explanation for current in the LRS of filament-based memristors. In a follow up study, Chang {\it et al.} refined the concept of including different conduction mechanisms~\cite{memristor_chang}. They defined the $i$--$v$ relation and the state change to directly model a Schottky emission~\cite{conduction_mechanism_survey} and tunnelling mechanism~\cite{simmons_tunnelling}. Both terms are weighted so that Chang's model behaves like a Schottky barrier in the HRS, and like a MIM junction in the LRS. The model depends on seven parameters. Yakopcic {\it et al.} proposed a memristor device model which was aimed at dissolving discrepancies between existing models~\cite{generalized_memristor_model}. While not rooted in physical principles, it offers more degrees of freedom in order to be fitted to a wide range of memristor data and is commonly referred to as {\it generalized model}~\cite{generalized_memristive_spice_model}. Here, the current can be modelled differently for the positive and negative regions of the input voltage and is specified by eleven fitting parameters. 

The evolution of memristor models introduced an ever-in-creasing number of parameters, and limitations in ordinary fitting algorithms became apparent. Recently, Yakopcic {\it et al.} proposed a model optimization approach based on parameter extraction~\cite{updated_generalized_model}. With it, they defined a memristor model, which can be interpreted as a generalization of Bakar {\it et al.}~\cite{abu_bakar_model}. In this approach, the $i$--$v$ equation is expressed as
\begin{equation}
	i = x\,h_1(v) + (1-x)\,h_2(v),
	\label{eq:iv_yakopcic_model}
\end{equation}
with $x \in [0, 1]$ being the normalized state variable. Here,
\begin{equation}
	h_k(v) = \begin{cases}
		\sigma v(t)\, , & \text{Ohm}\\
		\alpha (1-e^{-\beta v(t)})\, , & \text{Schottky}\\
		\gamma \sinh (\delta v(t))\, , & \text{MIM}\\
	\end{cases}
	\label{eq:hk_yakopcic_model}
\end{equation}
provides a generic placeholder for the conducting mechanism in the LRS ($k=1$) and HRS ($k=2$). In~\cite{updated_generalized_model}, it is reasoned that the $i$--$v$ relationships for the LRS and HRS can be deducted visually by looking at the curve shape of the pinched hysteresis loop (PHL) measurement data. Like the generalized model, this updated model uses $\dot{x}=g(v)f(x)$ to compute the state change, with $f(x)$ and $g(v)$ specifying the window function and internal dynamics, respectively. To conclude, it shall be noted that we have reviewed mainly an excerpt of electrical models, which showcase the evolution of universally applied mechanisms of conduction and state change. Various other electrical models have been proposed in the literature during the last decade, with varying levels of accuracy and for different purposes~\cite{understanding_memristive_switching}. In our findings, \eqref{eq:iv_yakopcic_model} can be understood as the current state of the art regarding electrical models and has been used to fit data to a wide range of memristors of varying types~\cite{updated_generalized_model}.
	
\subsection{Superstatistics}
The framework of superstatistics has been introduced by Beck and Cohen~\cite{superstatistics_2003}. It concerns systems driven away from equilibrium and exhibiting large spatiotemporal fluctuations of some intensive quantity, $\beta$, such as temperature, chemical potential, or energy dissipation. An illustration of this can be exemplified by Fig.~\ref{fig:gamma_noise_closeup}, wherein a system map is spatially divided into small cells (pixels in the figure) and the value of $\beta$ (color of the pixel) assigned to each cell is considered to be constant within the cell. For each of these local cells, the subsystem's state of energy for example can be described by the classical Boltzmann-Gibbs statistics. However, for larger scales, the whole system should be rather explained via a spatiotemporal average of the fluctuating parameter $\beta$. This results in a generalized Boltzmann factor being the integral
\begin{equation}
	B(E) = \int_0^\infty f(\beta) e^{-\beta E}d\beta\, ,
	\label{eq:generalized_boltzmann_factor}
\end{equation}
where $e^{-\beta E}$ defines the ordinary Boltzmann weight, $E$ is the effective energy in each cell and $f(\beta)$ expresses the probability distribution function (PDF) for  $\beta$. In~\eqref{eq:generalized_boltzmann_factor}, the whole system is described as a superposition of two statistics, $f(\beta)$ and $e^{-\beta E}$, which gives rise to the name superstatistics~\cite{superstatistics_applications_2004}. Eq.~\eqref{eq:generalized_boltzmann_factor} can also be interpreted as ({\it i}) the unconditional density of $E$ obtained from the conditional density  of $E$ at a given value of $\beta$, {\it i.e.}, $e^{-\beta E}$, and a marginal density of $\beta$ given by $f(\beta)$, or also as ({\it ii}) the Laplace transform of the function $f(\beta)$ giving $B(E)$. 

In principle, the function $f(\beta)$ should be determined a priori from the spatiotemporal dynamics of the entire system under study, which is unpractical in most situations. Since $f(\beta)$ describes the fluctuations of a positive real scalar random variable  in this study, distributions which are supported on the whole real axis, or distributions which are bounded --- like for example the normal distribution or the uniform distribution --- are unsuitable. Although one might find a broad selection of PDFs which are supported on the semi-infinite interval $[0, \infty)$, in this work, the gamma PDF will be explored. It is an appropriate choice for our system, owing to its versatility and flexibility; distributions like the Weibull, chi-square, Laplace, Maxwell-Boltzmann and other related densities can be obtained as special cases. The gamma distribution arises from the sum of $n$ independent Gaussian random variables with average $0$, which are squared and added. It can be understood as the distribution of a fluctuating environment with $n$ degrees of freedom~\cite{superstatistics_2003}. In its two-parameter form, the gamma PDF is written as
\begin{equation}
	f(\beta) = \frac{1}{b\Gamma(c)} \left(\frac{\beta}{b}\right)^{c-1} e^{-\frac{\beta}{b}}\,,
	\label{eq:generalized_boltzmann_gamma_0}
\end{equation}
where $\Gamma(c)$ is the gamma function and $b,c$ are positive constants, called {\it scale} and {\it shape} ($n=2c$) parameters. The mean and variance are given by
\begin{equation}
	\mathbb{E}(\beta) = \beta_0 = bc\, , \quad \mathrm{Var}(\beta) = b^2c\,.
    \label{eq:gamma_expected_variance}
\end{equation}

With $f(\beta)$ given by \eqref{eq:generalized_boltzmann_gamma_0}, the generalized Boltzmann factor in \eqref{eq:generalized_boltzmann_factor} can be formulated as
\begin{eqnarray}\nonumber
	B(E) & = &  (1+bE)^{-c}\\ \,
	& = & [1+(1-q)(-\beta_0E)]^{\frac{1}{1-q}}=e_q^{-\beta_0E}\, ,
	\label{eq:generalized_boltzmann_gamma_1}
\end{eqnarray}
where
\begin{equation}
	q=1+\frac1c\,
	\label{eq:q_from_c}
\end{equation}
indicates the {\it deformation} of the exponential~\cite{laplace_transforms_distributions, tsallis_q_exponential_definition}. One can verify that at the limit  $q \to 1$, $e_q(x)=e^x$. It shall be noted that \eqref{eq:generalized_boltzmann_gamma_0} requires $c$ to be positive and thus $q > 1$. However, a duality exists in which $q' = 2-q$, which allows to consider cases where $q < 1$~\cite{q_deformed_duality}. 

Superstatistics provide a natural way to extend Boltzmann-Gibbs statistics to a more general class of power-law distributed dynamics which can exhibit long-range interactions, metastability, or driving forces that keep the system out of equilibrium~\cite{superstatistics_applications_2004}. The literature provides numerous examples of observations which follow power-law distributions, namely chemical reactions between metals and chloride solutions exhibit power law behavior~\cite{power_law_chemical_reaction}, wind power persistence in Europe shows heavy tails for low- and high-velocity~\cite{superstatistics_wind_power}, frequency fluctuations in power grids can be characterized by superstatistics~\cite{power_grid_superstatistics}, and curved current-overpotential of Li-ion batteries have been modeled by $q$-deformed Butler--Volmer equations~\cite{deformed_bv_equations}. 

\section{Methodology}\label{sec:methodology}
\subsection{Experimental}
\begin{figure}
	\centering
	\subfloat[Voltage and current cycles over time]{
		\includegraphics[width=0.98\linewidth]{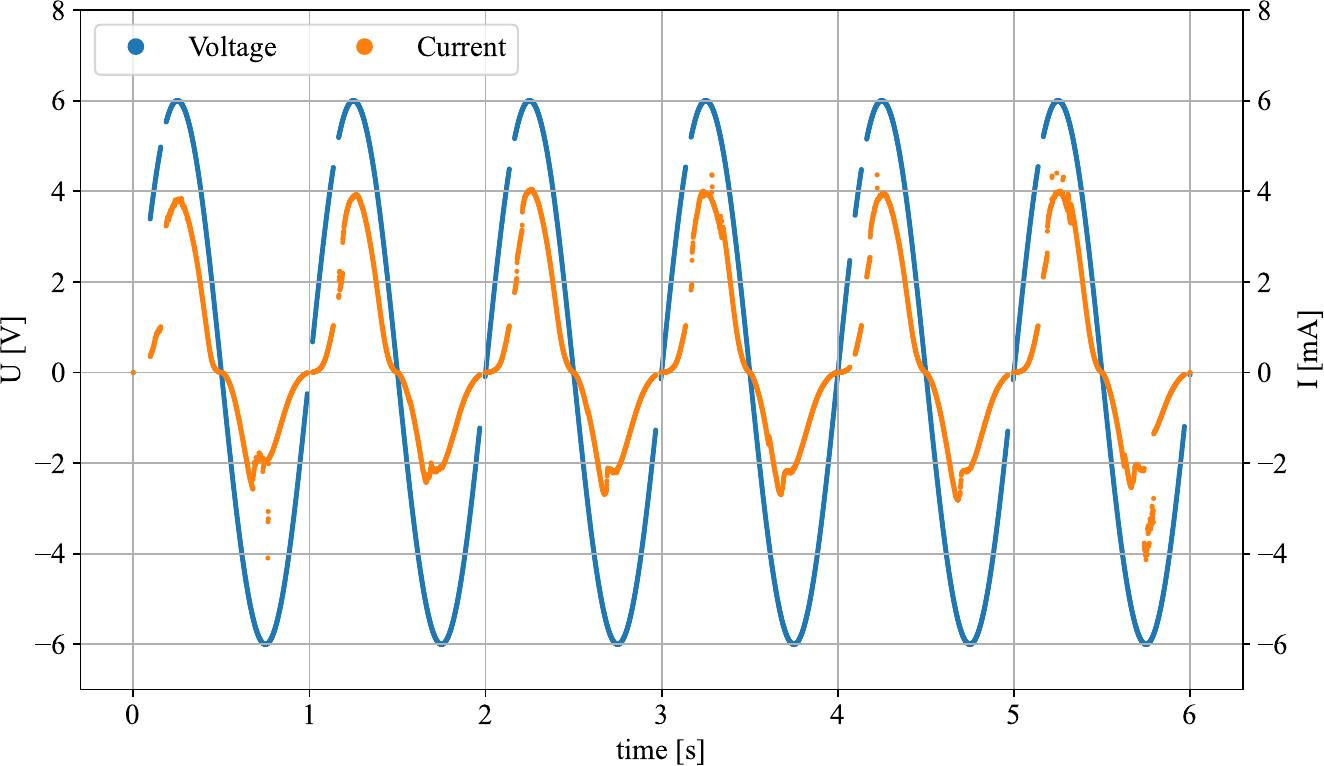}
		\label{fig:mem_measurements_vit}
	}\newline
	\subfloat[Separated $i$--$v$ cycles]{
		\includegraphics[height=0.46\linewidth]{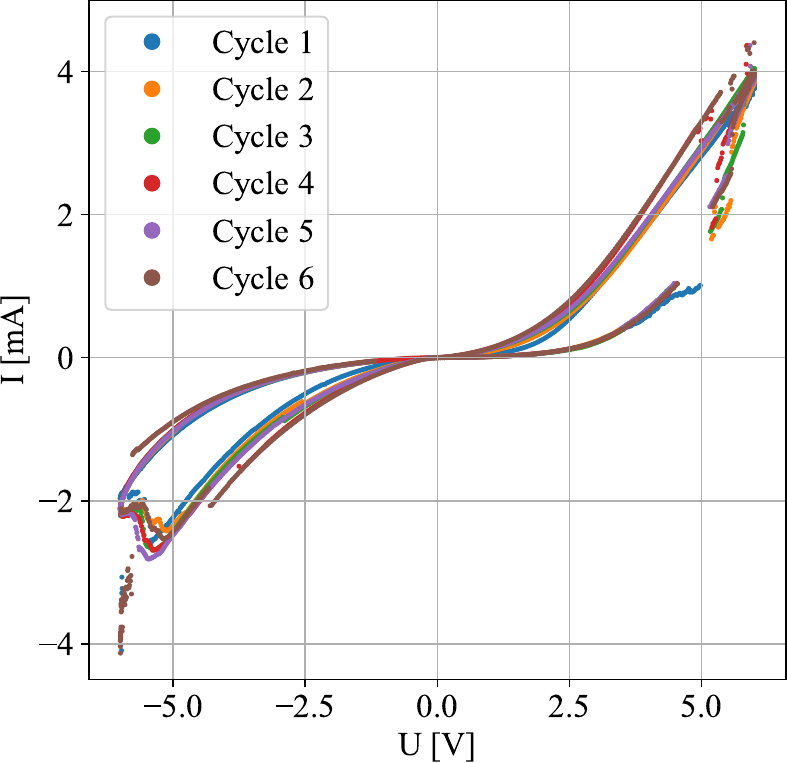}
		\label{fig:mem_measurements_iv}
	}
	\subfloat[Averaged $i$--$v$ cycle]{
		\includegraphics[height=0.46\linewidth]{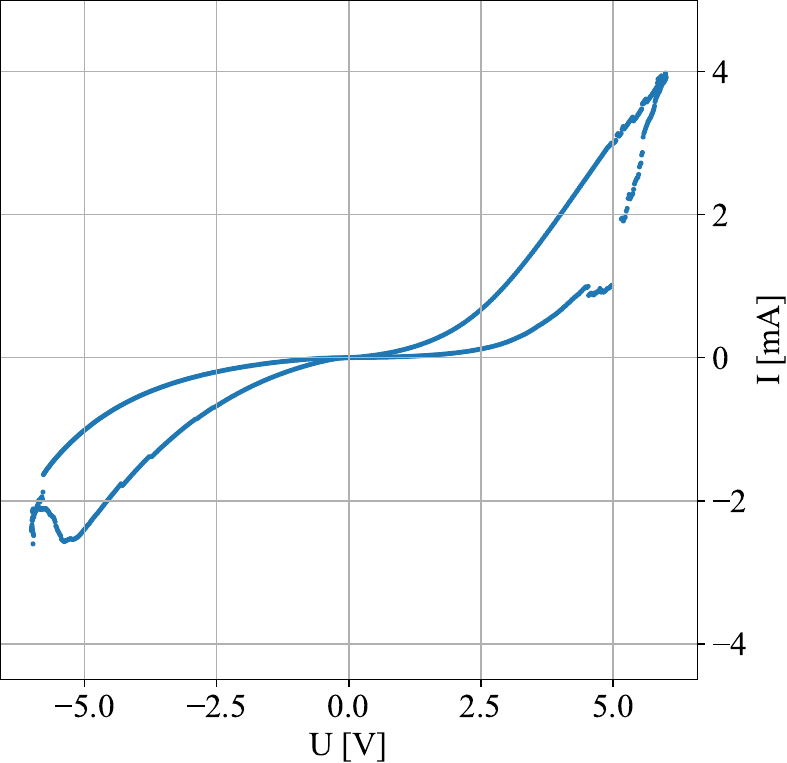}
		\label{fig:mem_measurements_iv_avg}
	}
	\caption{Measured memristor data}
\end{figure}

The device under test for this work is a Ag--Cu-based memristor which was fabricated  following the procedure of Yeon {\it et al.}~\cite{yeon_2020}. A schematic of the fabricated memristor can be seen in Fig.~\ref{fig:gamma_noise_closeup}. For the electrical measurements, a series of six sinusoidal voltage waveforms with 1 Hz frequency and 6 V amplitude were applied to the memristor using a BioLogic VSP-300 potentiostat workstation. The resulting current through the device was measured with a time resolution of 1~ms. The applied voltage signal, as well as the measured current over time can be seen in Fig.~\ref{fig:mem_measurements_vit}. The current--voltage cycles are presented in Fig.~\ref{fig:mem_measurements_iv}. An average cycle was calculated with those six sine waves, and is depicted in Fig.~\ref{fig:mem_measurements_iv_avg}. As the typical endurance of filament-based memristors is in the order of $10^6$ cycles~\cite{memristor_physics_part_1, memristor_filament_degradation}, the current state of degradation for the device under test was considered to be effectively static for the six recorded $i$--$v$ responses.

\subsection{Initial model evaluation}
In order to develop a memristor model using superstatistical methods, an initial evaluation of currently used models was conducted. The best performing model was then taken and used as the basis for further development. For this initial selection, the models were fitted to one, averaged $i$--$v$ cycle, as depicted in Fig.~\ref{fig:mem_measurements_iv_avg}. In total, six models were considered for the initial evaluation, including the model by Chang {\it et al.}~\cite{memristor_chang}, the generalized model by Yakopcic {\it et al.}~\cite{generalized_memristor_model}, the model by Bakar {\it et al.}~\cite{abu_bakar_model}, as well as three variants of the updated model by Yakopcic {\it et al.}~\cite{updated_generalized_model} as specified by \eqref{eq:iv_yakopcic_model}. The latter three variants were defined by the selection of $h_k(v)\;(k=1,2)$ in \eqref{eq:hk_yakopcic_model} as follows:
\begin{equation}
	h_1(v)=\sigma v \,, \quad h_2(v)=\alpha(1-e^{-\beta v})
\end{equation}
for the so-called Yakopcic OS model (Ohm - Schottky),
\begin{equation}
	h_1(v)=\gamma\sinh(\delta v)\,, \quad h_2(v)=\alpha(1-e^{-\beta v})
\end{equation}
for the Yakopcic MS model (MIM - Schottky), and
\begin{equation}
	h_1(v)=\gamma\sinh(\delta v)\,, \quad h_2(v)=\gamma\sinh(\delta v)
	\label{eq:iv_yakopcic_model_mim}
\end{equation}
for the Yakopcic MM model (MIM-MIM). The models were chosen as they showcase the evolution of memristor understanding, and exhibit a trend of increasing complexity.

\subsection{Model fitting}
In the literature, the common metric of quality for memristor models has not changed for the most part, as the measured PHL still forms the main reference for those models to be compared against. Challenges with fitting ever-more complex models to the data are an emerging trend, which are addressed in \cite{updated_generalized_model}. Fine-tuning initial fitting parameters or setting parameter boundaries are a few examples for how operators can achieve stable fitting results; however, it is the opinion of the authors that enough data, a suitable model and a capable fitting algorithm should be sufficient to achieve a stable, optimal result without further human intervention. In this work, a heuristics-based, stochastic global fitting procedure has been implemented, which is based on the basin-hopping algorithm~\cite{Wales_basinhopping, Luo_basinhopping}. A detailed explanation of the method can be found in Section 3.2 in~\cite{konlechner_memristor}.

\section{Results and discussion}\label{sec:rnd}

\begin{figure}
	\includegraphics[width=0.95\linewidth]{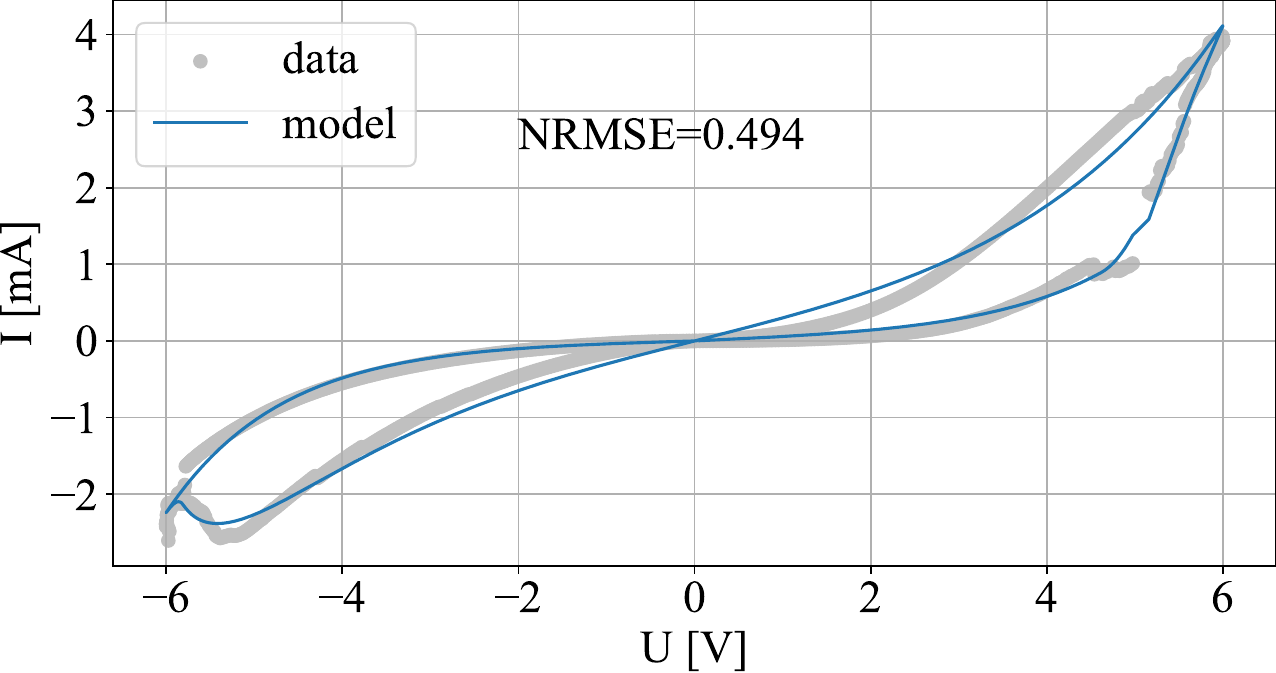}
	\caption{Yakopcic MM model fitted to measured $i$--$v$ response}
	\label{fig:first_sel_yakopcic_mm}
\end{figure}

Our numerical findings suggest that the Yakopcic MM model performed the best on the given data, with the lowest normalized root-mean-square error (NRMSE). In Fig.~\ref{fig:first_sel_yakopcic_mm}, the resulting $i$--$v$ curve of the Yakopcic MM model is shown, plotted against the data. Table~\ref{table:fitting_results}  shows all tested models with their optimized parameters and final NRMSEs. When inspecting the measured $i$--$v$ relationship in Fig.~\ref{fig:mem_measurements_iv_avg}, resemblances of a hyperbolic sine shape can be seen for both the LRS and HRS. The result of this initial fitting therefore also makes intuitive sense. For the next step, the Yakopcic MM model was used as the basis for building the $q$-deformed model.
	
\subsection{q-deformed models development and validation}
As seen in~\eqref{eq:iv_yakopcic_model_mim}, the Yakopcic MM model uses a parameterized hypberbolic sine function for both the LRS and HRS of~\eqref{eq:iv_yakopcic_model}, with $\gamma$ and $\delta$ being the parameters to be fitted~\cite{memristive_switching_mechanism, memristor_chang}. To develop a superstatistical approach, the current equation for ionic conduction is to be studied. This mechanism can be interpreted from a transition state formalism point of view, and can be generally expressed as
\begin{eqnarray}
	J_\mathrm{ionic} & \propto & v_Dr \cdot\exp\left(-\frac{\Delta G^{\neq}}{k_BT}\right)\nonumber\\
	&& \times \left\{\exp\left(\frac{re}{2k_BT}E\right) - \exp \left( -\frac{re}{2k_BT}E\right) \right\},
	\label{eq:ionic_current_approx}
\end{eqnarray}
where $v_D$ is the Debye frequency, $r$ is the jump distance, $\Delta G^{\neq}$ stands for the free activation enthalpy, $k_B$ is the Boltzmann constant, $T$ denotes the temperature of the system, $e$ is the electron charge, and $E$ represents the electric field~\cite{cation-based_resistance_change_memory, electrochemical_metallization_memories}. With the simplifying assumption that $E \propto v$, and by setting
\begin{equation}
	\gamma = 2\cdot v_Dr\cdot\exp\left(-\frac{\Delta G^{\neq}}{k_BT}\right)\, , \quad \delta = \frac{re}{2k_BT}\,,
	\label{eq:gamma_delta_definition}
\end{equation}
we rewrite $J_\mathrm{ionic}$ as
\begin{equation}
	J_\mathrm{ionic} \propto \frac{\gamma}{2} \cdot \left( e^{\delta v} - e^{-\delta v} \right) = \gamma\sinh(\delta v) \,,
	\label{eq:ionic_current_to_yakopcic_iv}
\end{equation}
therefore utilizing ionic conduction as the mechanism for~\eqref{eq:iv_yakopcic_model_mim}, a process which follows Boltzmann-Gibbs statistics.
	
Filament-type memristors rely on ion migration as the basic mechanic for state change~\cite{williams_memristor_found, joglekar_window_function, biolek_window_function, prodromakis_window_function, memristor_chang,jo_programmable_resistance_switching}, which means that the change of state inherently causes an ionic current in the device. Even after the memristor is in the LRS, the filament has been observed and simulated to grow thicker with changing current compliance~\cite{kmc_simulation_of_hfo2,modeling_universal_set-reseet_characteristics_of_rram}. This hints at ions still migrating after the conduction channel has been formed. As the device ages over many set/reset cycles, it experiences degradation due to parasitic filament formation~\cite{memristor_physics_part_1, memristor_filament_degradation}. This ageing can be explained by stray ions permeating the entire switching layer of the memristor, which slowly accumulate at one electrode and form parasitic filaments. This again is a form of ionic current. The literature thus provides numerous examples as to why ionic conduction can be at least partially attributed to the $i$--$v$ characteristics measured in memristive devices, and justifies this approach for further model development.
	
With~\eqref{eq:iv_yakopcic_model_mim} now established through \eqref{eq:ionic_current_approx} and \eqref{eq:gamma_delta_definition}, the updated $i$--$v$ relationship of the $q$-deformed model can be defined by using the $q$-deformed exponential~\eqref{eq:generalized_boltzmann_gamma_1} in place of the exponential functions in~\eqref{eq:ionic_current_to_yakopcic_iv}. With the $q$-deformed hyperbolic sine being expressed as
\begin{equation}
	\sinh_q(x) = \frac{e_q(x) - e_q(-x)}{2}\,,
\end{equation}
the complete $i$--$v$ relationship of the $q$-deformed model can now be defined as
\begin{equation}
	i=\gamma_1x\sinh_q(\delta_1 v) +\gamma_2(1-x)\sinh_q(\delta_2 v)\, ,
	\label{eq:iv_q_deformed_mm_model}
\end{equation}
where $\gamma_1$, $\delta_1$, $\gamma_2$, $\delta_2$, and $q$ are fitting parameters. For reference within the rest of this work, this model will be denoted as $q$-deformed MM model.
	
Intuitively, the $q$-deformed MM model can be understood as follows: The spatial distribution of mobile dopants in the substrate is not homogeneous, but follows some distribution function. These inhomogeneities have been observed and are attributed to nanoparticle formation due to Rayleigh instability~\cite{optical_properties_co_sputtering, cantor-microstructore-si-ag-nanocomposites}. The resulting non-uniformity describes a system with stationary, non-equilibrium states, the type of system which is explained by superstatistics. If a voltage potential is applied to the memristor switching layer, an electric field forms, which generates cations and encourages them to move. The electric field depends on the applied voltage potential, as well as the thickness of the dielectric. Since the mobile dopants are not uniformly distributed, the thickness of the dielectric is not uniform either. By using the electric field as the intensive parameter $\beta$, and by assuming that the distribution of $\beta$ can be described by the gamma PDF, the generalized Boltzmann factor as shown in \eqref{eq:generalized_boltzmann_gamma_1} is obtained. Here, the newly acquired parameter $q$ can be interpreted as a measure of the non-uniformity of dopants in the switching layer.

\begin{figure}
	\centering
	\subfloat[$q=0.9$.]{
		\includegraphics[width=0.40\linewidth]{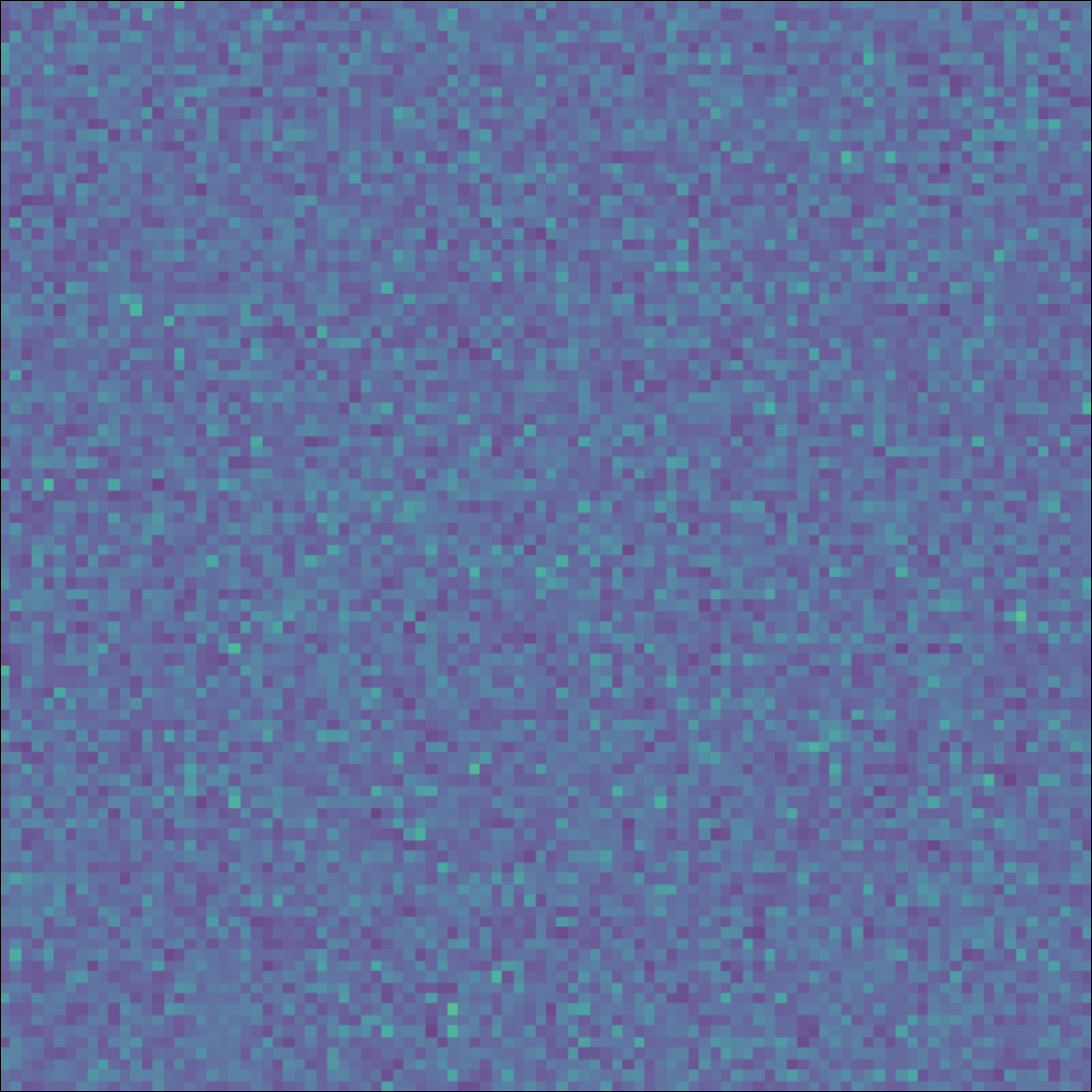}
		\label{fig:gamma_noise_0.9}
	}
	\hspace*{2em}
	\subfloat[$q=0.5$.]{
		\includegraphics[width=0.40\linewidth]{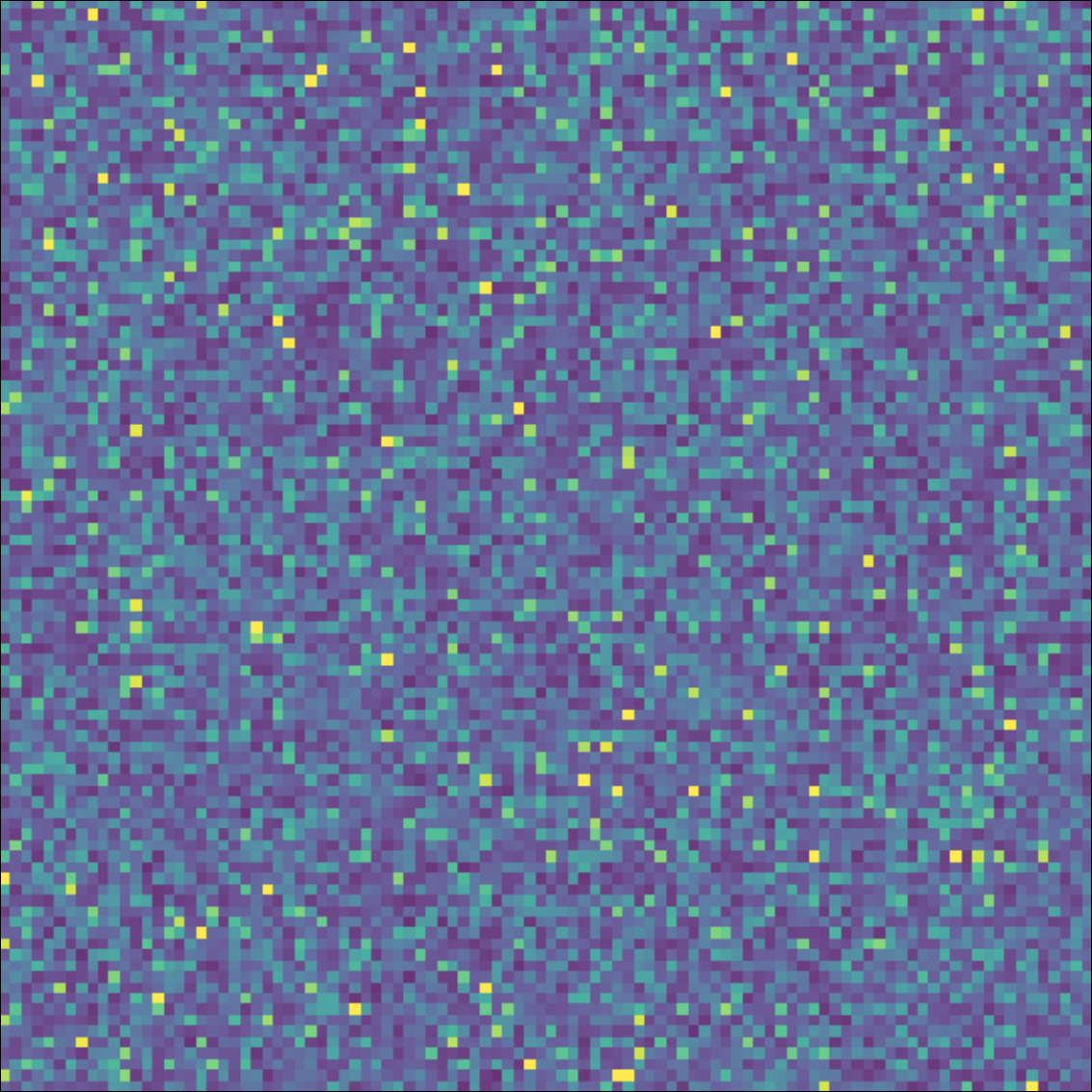}
		\label{fig:gamma_noise_0.5}
	}
	\caption{Normalized gamma noise samples for different $q$}
	\label{fig:gamma_q_comparison}
\end{figure}

Fig.~\ref{fig:gamma_q_comparison} illustrates this relationship for different values of $q$. As shown in \eqref{eq:q_from_c}, only the shape $c$ of the gamma PDF depends on $q$. Here, the duality of~\cite{q_deformed_duality} is used to rewrite $c$ as
\begin{equation}
	c = \frac{1}{(2-q)-1}
\end{equation}	
to consider values of $q<1$. In \eqref{eq:gamma_expected_variance}, one can see that the scale $b$ of the gamma PDF is independent of $q$; it was therefore defined as $b = 1/c$ to normalize the mean value $\beta_0 = bc = 1$. As can be seen in Fig. \ref{fig:gamma_q_comparison}, the spatial inhomogeneities for $q=0.9$ are lower than for $q=0.5$.

\begin{figure}[ht]
	\centering
	\subfloat[$q$-deformed MM model. $q$-value: $0.726$]{
		\includegraphics[width=0.91\linewidth]{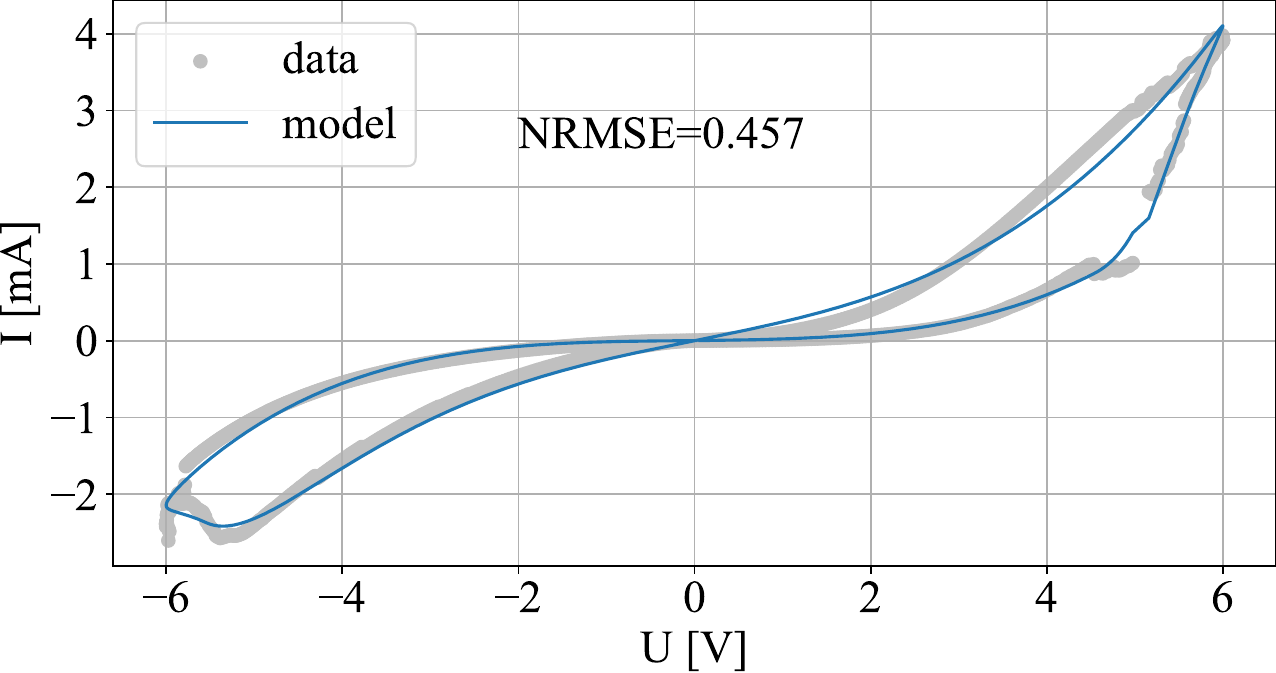}
		\label{fig:second_q_deformed_mm}
	}\newline
	\subfloat[$q$-deformed MM state model. $q$-value: $0.496$]{
		\includegraphics[width=0.91\linewidth]{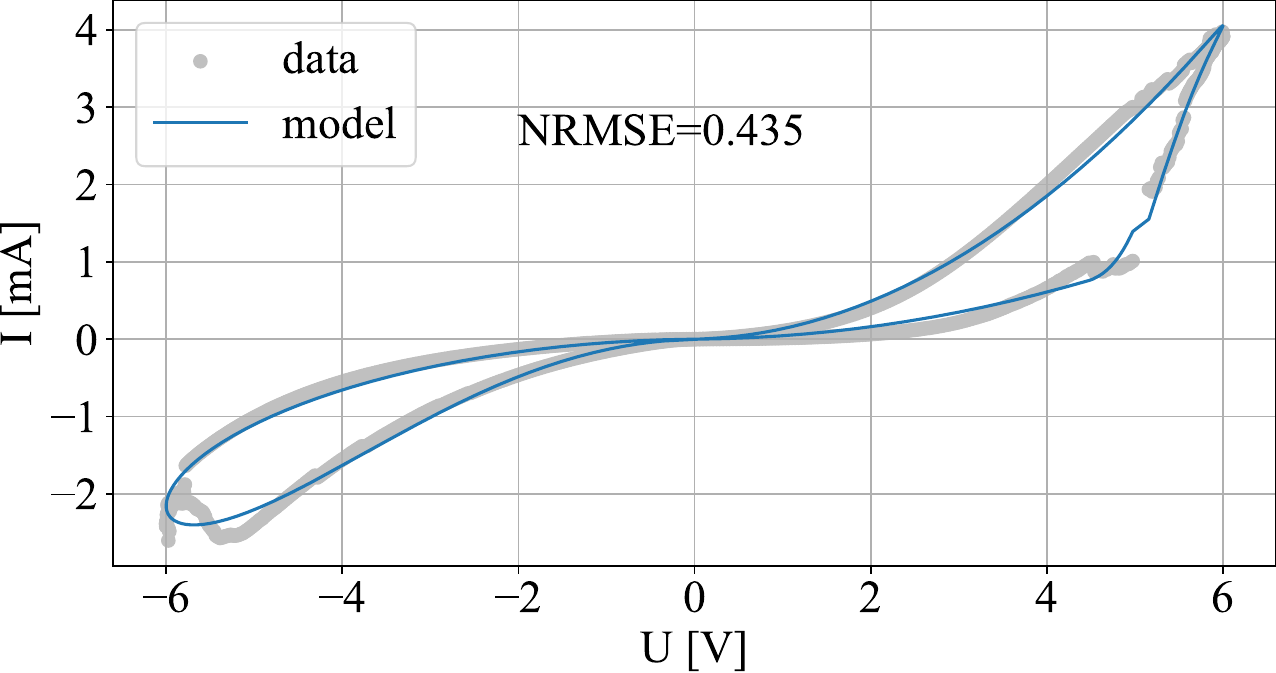}
		\label{fig:second_q_deformed_mms}
	}\newline
	\subfloat[$q$-deformed M state model. $q$-value: $0.499$]{
		\includegraphics[width=0.91\linewidth]{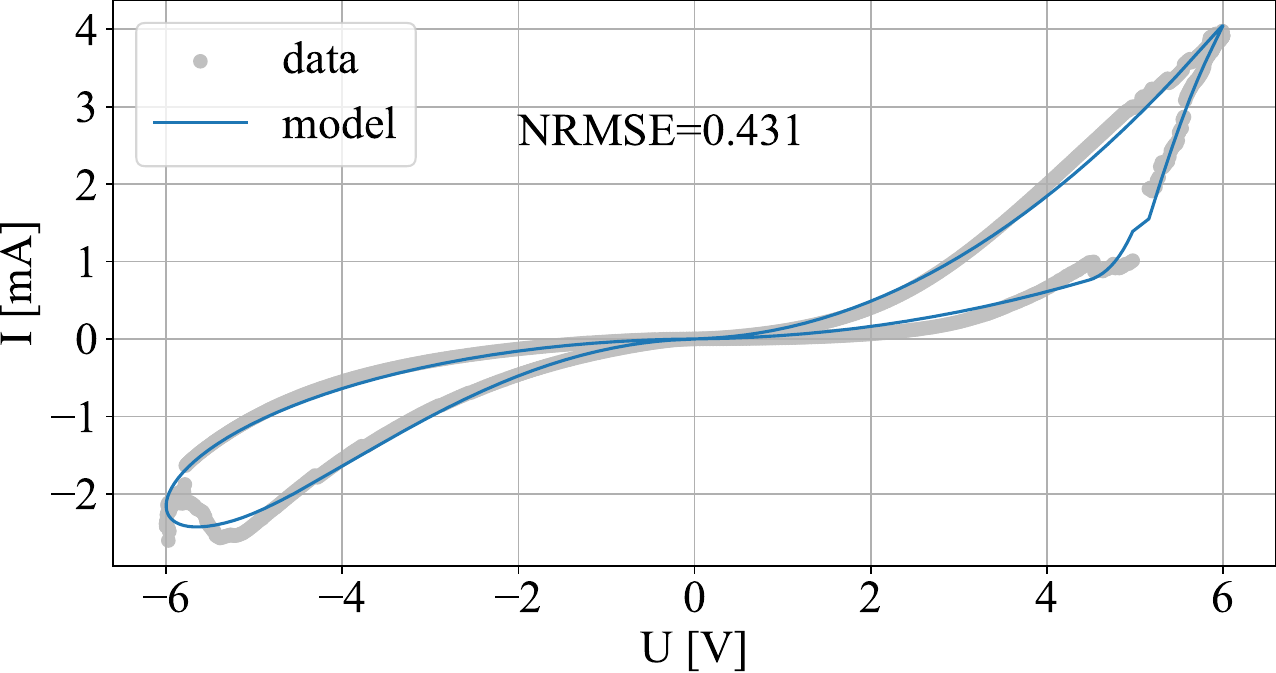}
		\label{fig:second_q_deformed_ms}
	}
    \caption{$q$-deformed models fitted to measured $i$--$v$ response}
\end{figure}

To test the performance of the $q$-deformed MM model, it was fitted to the same, averaged PHL as the previous models. In Fig.~\ref{fig:second_q_deformed_mm}, the resulting $i$--$v$ curve of this model is shown, plotted against the data. The result shows an NRMSE of 0.457 for the $q$-deformed MM model, which constitutes an improvement of about 8\% over the baseline model.
	
Since ionic current is intrinsically tied with state change in filament-based memristors, the possibility of introducing the $q$-parameter to the state change equation was explored next. The Yakopcic MM model uses a closely related version of the state change equations for the generalized model~\cite{generalized_memristor_model} with $g(v)$ describing the state change and $f(x)$ defining a window function for non-linear ion movement. With this assumption, the state change in the $q$-deformed MM model would also be affected by dopant inhomogeneities. Thus, $g(v)$ was updated by replacing all exponentials with the $q$-deformed exponential, resulting in $g_q(v(t))$. The updated state change equation is now defined as
\begin{equation}
    \dot{x} = g_q(v(t)) f(x)\, .
    \label{eq:state_equation_mms_model}
\end{equation}
For the rest of this work, this model will be denoted as $q$-deformed MM state model, which utilizes~\eqref{eq:iv_q_deformed_mm_model} as the current equation and~\eqref{eq:state_equation_mms_model} as the state-change equation.
	
In Fig.~\ref{fig:second_q_deformed_mms}, the result of the $q$-deformed MM state model is plotted against the data. The results show that with a NRMSE of 0.435, another small improvement could be achieved with the $q$-deformed MM state model, which performs about 5\% better than the $q$-deformed MM model on the averaged PHL data. Visually, it is apparent that the $q$-deformed MM state model characterizes the hyperbolic sine shape in the LRS of the PHL more accurately.
	
An interesting observation can be made when studying the resulting final parameters. The fitting algorithm naturally eliminated the complete second term $h_2(v)$ in \eqref{eq:iv_yakopcic_model}, simplifying the $i$--$v$ relationship to
\begin{equation}
	i = \gamma x\sinh_q(\delta v) \, .
	\label{eq:iv_q_mstate_model}
\end{equation}
Following this observation, a simplified model with~\eqref{eq:iv_q_mstate_model} as the $i$--$v$ relationship was implemented--this model will be denoted as $q$-deformed M state model.

The fitting results of the $q$-deformed M state model can be seen in Fig.~\ref{fig:second_q_deformed_ms}. With a NRMSE of 0.431, the $q$-deformed M state model performs comparably to the $q$-deformed MM state model, although having less fitting parameters. This simplification however comes at the cost of state transition accuracy of the model. As can be seen in the bottom left corner in Figs.~\ref{fig:second_q_deformed_mms} and~\ref{fig:second_q_deformed_ms}, the complex transition shape from LRS to HRS in the PHL can not be accurately simulated by the models.

All three tested models exhibit a value of {$q < 1$}, which indicates a divergence from the baseline exponential-based model for fitting to this dataset. Visually, this {\it deformation} can be seen in the high resistance region of the measured data in Figs.~\ref{fig:second_q_deformed_mm}--\ref{fig:second_q_deformed_ms}, where the $q$-deformed models more accurately assume the curve shape over the baseline model.

\subsection{Further analysis}
The $q$-deformed models were developed with the premise that they are able to express internal device inhomogeneities from the given data. This assumption leads to a few theories, which are tested in this section. This work universally assumes the gamma PDF to explain the non-uniformity of mobile dopants in memristive devices. It can be shown that for $n$ independent and identically distributed variables $\{x_1, x_2, ..., x_n\}$, where
\begin{equation}
	x_i \sim\mathrm{Gamma}(k, \theta)\, , \quad i \in \{1, 2, ..., n \}\, ,
\end{equation}
the average of those variables is expressed as
\begin{equation}
	\langle x_n\rangle \sim \mathrm{Gamma}(k n, \theta/n)\, .
\end{equation}	Since the variance of the gamma PDF is defined as
\begin{equation}
	\mathrm{Var}(x) = k \theta^2\, ,
\end{equation}
the variance of the average becomes
\begin{equation}
	\mathrm{Var}(\langle x_n\rangle) = kn \left(\frac{\theta}{n}\right)^2 = k\frac{\theta^2}{n}\, .
\end{equation}
Thus, the variance of the average $\langle x_n\rangle$ gets smaller with an increasing number of variables. With this characteristic, it is assumed that by averaging an increasing number of single measured $i$--$v$ cycles, device inhomogeneities become less and less pronounced in the resulting PHL. Hence, $q$-deformed models should show a larger improvement over the baseline model when fitted to single cycles, as opposed to the average of multiple cycles.
	
To test this theory, the complete measured $i$--$v$ signal from the memristor was split into six separate cycles, as seen in Fig.~\ref{fig:mem_measurements_iv}, and from then on treated as single, independent $i$--$v$ datasets. By following the binomial coefficients and Pascal's triangle, subsets were generated, each containing between 1 and 6 single $i$--$v$ cycles. Table~\ref{table:cycles_subsets_combinations} shows the number of generated subsets for each possible size of the subset. With this combinatorial approach, 63 subsets were generated in total.
	
\begin{table}[h!]
    \centering
    \begin{tabular}{c c} 
        \hline\hline
        \textrm{Number of subsets} & \textrm{Size of subset (number datasets)} \\ [0.5ex] 
        \hline 
        6 & 1\\
        15 & 2\\
        20 & 3\\
        15 & 4\\
        6 & 5\\
        1 & 6\\  
        \hline\hline
    \end{tabular}
    \caption{Data subset combinations}
    \label{table:cycles_subsets_combinations}
\end{table}

As the next step, the $q$-deformed MM model, the $q$-deformed MM state model, the $q$-deformed M state model, and the Yakopcic MM model were each fitted to all 63 subsets. Afterwards, for each $k \in \{1, 2, ..., 6\}$, the $\mathrm{NRMSE}_{k,\mathrm{avg}}$ was calculated according to
\begin{equation}
	\mathrm{NRMSE}_{k,\mathrm{avg}} = \frac{\sum_{i=1}^n \mathrm{NRMSE}_{k,i}}{n}, \, n = \frac{6!}{k!(6-k)!}\, .
\end{equation}
Here, $\mathrm{NRMSE}_{k,i}$ is the NRMSE for the fitting run to subset $i \in \{1, 2, ..., n\}$ of size $k$. With this approach, $k$ can be understood as the {\it averageness} of the data --- the higher $k$, the more datasets are considered during fitting, and the more device inhomogeneities are averaged out. Following this understanding, $\mathrm{NRMSE}_{k,\mathrm{avg}}$ is a measure for how well the model performs (on average) to subsets of size $k$. Although $n$ does not directly affect the result, a larger $n$ reduces the variance of $\mathrm{NRMSE}_{k,\mathrm{avg}}$ for a given $k$, thus providing a statistically more relevant result.
	
An example is given to make the procedure more explicit. There are 15 subsets of size 4 ($n=15, k=4$). Each subset contains one of the 15 possible combinations of 4 single-cycle $i$--$v$ curves. The model is fitted to all 4 $i$--$v$ curves of one subset at once, thus generating 15 NRMSEs for each model ($\mathrm{NRMSE}_{4,i},$ $i \in \{1, 2, ..., 15\}$). Those 15 NRMSEs are then summed up and divided by 15, which results in an average NRMSE of this model for the subsets of size 4 ($\mathrm{NRMSE}_{4,\mathrm{avg}}$).

\begin{figure}[!h]
	\subfloat[Average performance of $q$-deformed models compared to Yakopcic MM model over all subset combinations]{
		\includegraphics[width=0.93\linewidth]{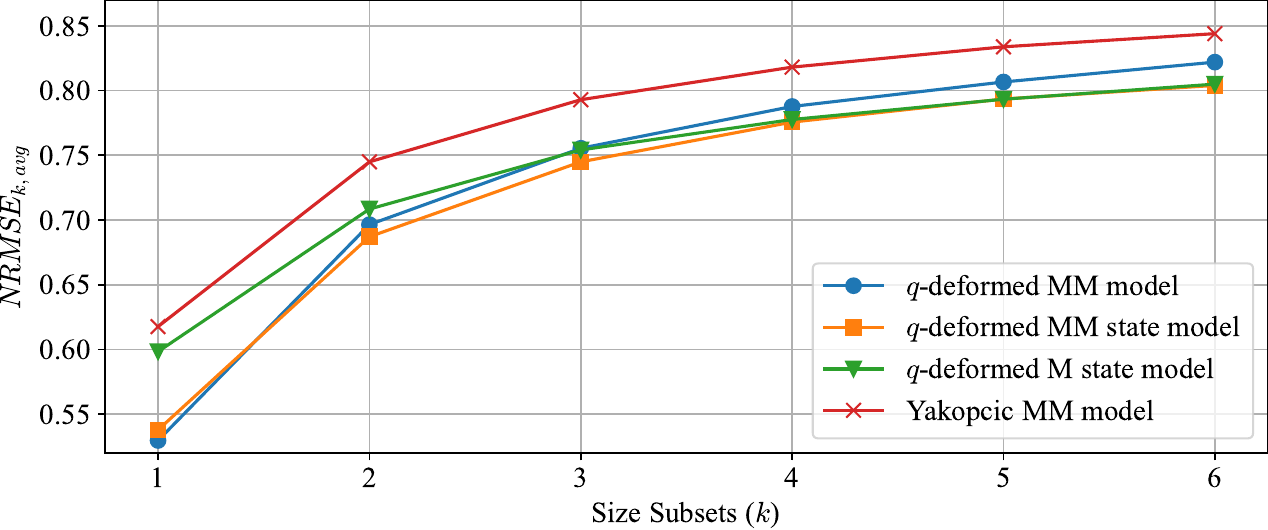}
		\label{fig:q_analysis_nrmse}
	}\newline
	\subfloat[Absolute improvement of $q$-deformed models compared to Yakopcic MM model over all subset combinations]{
		\includegraphics[width=0.93\linewidth]{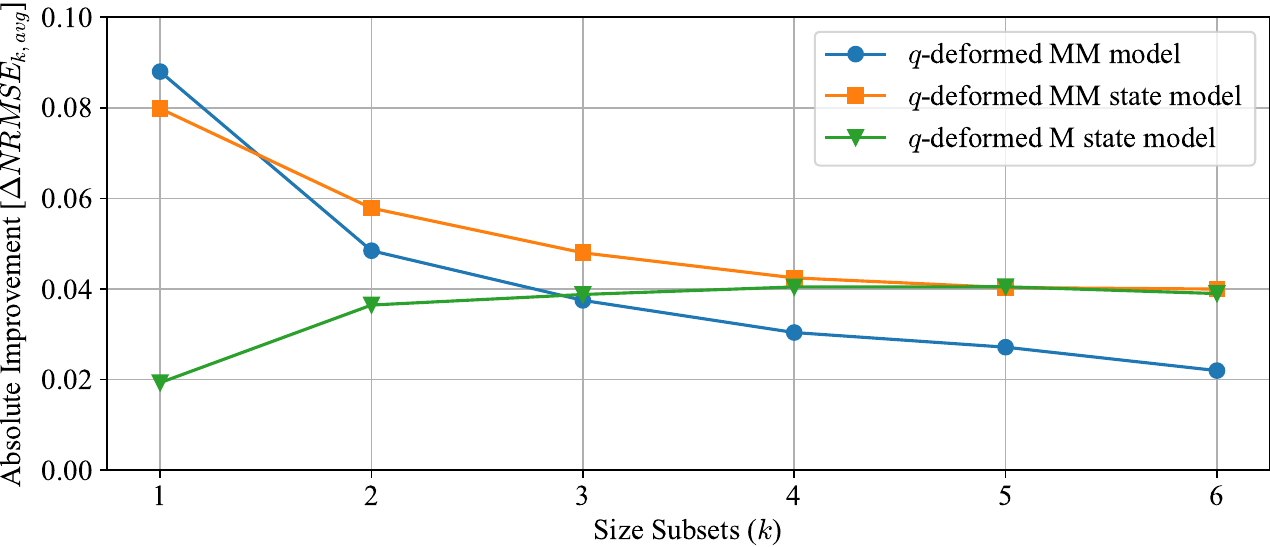}
		\label{fig:q_analysis_abs_improvement}
	}\newline
	\subfloat[Relative improvement of $q$-deformed models compared to Yakopcic MM model over all subset combinations]{
		\includegraphics[width=0.93\linewidth]{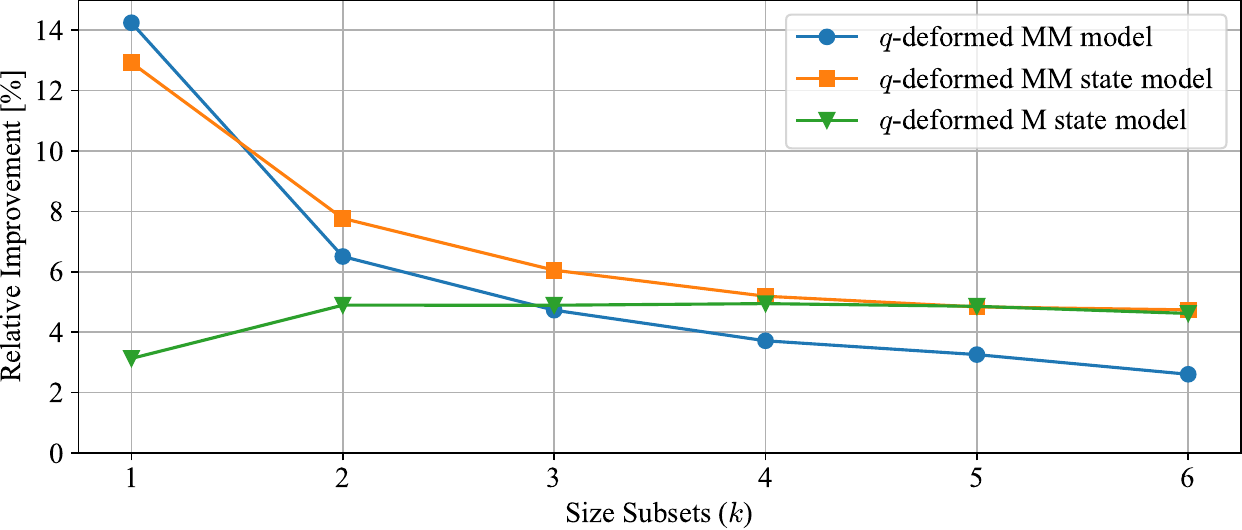}
		\label{fig:q_analysis_rel_improvement}
	}
	\caption{Analysis of $q$-deformed models with $i$--$v$ cycle subset combinations}
\end{figure}

Fig.~\ref{fig:q_analysis_nrmse} shows the $\mathrm{NRMSE}_{k,\mathrm{avg}}$ plotted over the subset size $k$ for each fitted model. What becomes apparent right away is that the average performance of all tested models is significantly better for single-cycle subsets, and gets worse for larger subsets. This observation agrees with model fitting expectations. Smaller subsets allow the models to assume noise in the data, thus showing overfitting tendencies. This effect is mitigated with a growing size of the subsets, which results in a more general solution for the model. This assumption is confirmed by the observation of $\mathrm{NRMSE}_{k,\mathrm{avg}}$, which flattens off for a growing $k$. Fig.~\ref{fig:q_analysis_nrmse} also shows that all $q$-deformed models perform on average significantly better than the baseline Yakopcic MM model for all subset sizes $k$.
	
Figs.~\ref{fig:q_analysis_abs_improvement} and~\ref{fig:q_analysis_rel_improvement} depict the absolute and relative improvements of the $q$-deformed models over the baseline Yakopcic MM model, plotted over the subset sizes $k$. The absolute improvement was calculated as the difference $\Delta\mathrm{NRMSE}_{k,\mathrm{avg}}$ between each $q$-deformed model and the baseline model. To obtain the relative improvement, $\Delta\mathrm{NRMSE}_{k,\mathrm{avg}}$ was normalized with
\begin{equation}
	\text{Relative Improvement} = \frac{\Delta\mathrm{NRMSE}_{k,\mathrm{avg}}}{\mathrm{NRMSE}_{k,\mathrm{avg}|\mathrm{Yakopcic}}}
\end{equation}
and is shown as a percentage. Figs.~\ref{fig:q_analysis_abs_improvement} and~\ref{fig:q_analysis_rel_improvement} confirm the main premise of the $q$-deformed models. For smaller subset sizes $k$, the $q$-deformed models show a significantly higher improvement than for larger subset sizes. The $q$-deformed MM model and the $q$-deformed MM state model perform around $14\%$ and $13\%$ better for single cycles, which drops to around $3\%$ and $5\%$ for a growing $k$. It is also clear that the $q$-deformed M state model does not follow this common trend. Although the improvement is comparable to the $q$-deformed MM state model for a $k\geq 5$, it performs significantly worse for smaller subset sizes.
	
\section{Conclusion and outlook}\label{sec:conclusion}
The main results in this work demonstrate that the superstatistical approach was successfully implemented to model memristors. The developed $q$-deformed models performed 4--14\% better than the baseline model for the various conducted tests. However, this statement is made purely from the viewpoint of model fitting, {\it i.e.}, only the ability of the model to accurately assume the data is analyzed. Since a new parameter is introduced with $q$, and forms a generalization of the baseline model (which is retrieved with $q=1$), it is not far fetched to achieve a better fitting result. Underpinning such a model with a physical foundation is an entirely different challenge.

Although conclusive statements about the underlying physics are difficult to make, the conducted experiments show definite hints which suggest that the main premises made in this work hold true. The obtained results demonstrate that the $q$-deformed models show a significantly larger improvement for single $i$--$v$ cycles, and get progressively closer to the baseline model with a growing cycle size. This observation leads to the conclusion that $q$-deformed models indeed better characterize device inhomogeneities. Another potential clue for the validity of this approach is the fact that the $q$-deformed MM state model, where the $q$-deformation parameter was also introduced into the state change formula, leads to an overall better performance than the $q$-deformed MM model. Since the main explanation for the state change in cation-based memristors is the growth and rupture of filaments, the inherent ionic current in this state change would be equally affected by the device inhomogeneities explained by the $q$-parameter. The improved performance of the $q$-deformed MM state model gives some proof to this statement.
		
The fitting process automatically eliminated two parameters of the $q$-deformed MM state model, giving rise to the reduced, $q$-deformed M state model. This reduction of complexity can also be observed in Figs.~\ref{fig:q_analysis_abs_improvement} and~\ref{fig:q_analysis_rel_improvement}, where the performance of both models converge for a larger $k$. Since this effect occurs purely for multiple $i$--$v$ cycles and averaged data, the reduction in model complexity seems to be a consequence of more general dataset characteristics, and less of single-cycle noise. The memristor which was used for this work underwent several experiments before the $i$--$v$ measurements were recorded. The age and general degradation of the device were therefore unknown before measurements were taken, and could be one possible reason for the observed reduction in complexity for the $q$-deformed MM state model.

The underlying model from Yakopcic {\it et al.} is rather general and has been tested in multiple scenarios with different kinds of memristors~\cite{generalized_memristor_model,updated_generalized_model,generalized_memristive_spice_model}. Since this model forms a special case of the $q$-deformed models introduced in this work, we reason that the $q$-deformed models have more general applicability to different kinds of memristors as well. However, it shall be noted here that this reasoning is purely deductive and still needs to be verified.

Proceeding this work, there are several possibilities for further investigation. As the most straightforward next step, the resulting $q$-deformed models need to be verified for different $i$--$v$ responses of different types of memristors. Additionally, the evolution of the $q$-parameter for $i$--$v$ responses of a progressively ageing memristor could give valuable insights about the correlation of $q$ with the general state of degradation of the device.

\section*{Acknowledgments}
{\bf Dmitry Yudin} acknowledges the support from the Russian Science Foundation Project No. 22-11-00074.
 
\appendix
\renewcommand{\thefigure}{A\arabic{figure}}
\setcounter{figure}{0}
\renewcommand{\thetable}{A\arabic{table}}
\setcounter{table}{0}

\section*{Additional Material}

\begin{table*}
    \centering
    \begin{tabular}{l l c} 
        \hline\hline
        \textrm{Model} & \textrm{Parameters} & \textrm{NRMSE} \\ [0.5ex] 
        \hline 
        Chang \cite{memristor_chang} & $\alpha=0.0$, $\beta=1.764$, $\gamma=0.132$, $\delta=0.575$, $\lambda=3.525$, $\eta_1=0.121$, $\eta_2=0.184$ & 0.649\\
        Generalized \cite{generalized_memristor_model} & $\alpha_p=10.227$, $\alpha_n=8.768$, $x_p=0.586$, $x_n=0.326$, $A_p=0.055$, $A_n=0.043$, $V_p=0.0$,  & 0.533\\
         & $V_n=4.24$, $a_1=0.686$, $a_2=0.612$, $b=0.451$ & \\
        Bakar \cite{abu_bakar_model} & $\alpha_p=621.853$, $\alpha_n=52.076$, $x_p=0.001$, $x_n=0.999$, $A_p=0.462$, $A_n=0.001$, $V_p=4.977$, & 0.578\\
        & $V_n=0.634$, $\alpha=9413.917$, $\beta=0.002$, $\gamma=0.05$, $\delta=0.753$ & \\ 
        Yakopcic OS \cite{updated_generalized_model} & $x_p=0.944$, $x_n=0.258$, $A_p=0.089$, $A_n=0.038$, $V_p=2.444$, $V_n=0.0$, $\sigma=0.568$, & 1.242\\
        & $\alpha=0.006$, $\beta=0.958$ & \\
        Yakopcic MS \cite{updated_generalized_model} & $x_p=0.621$, $x_n=0.788$, $A_p=0.07$, $A_n=0.05$, $V_p=0.0$, $V_n=0.0$, $\alpha=0.029$, $\beta=0.726$, & 0.506\\
        & $\gamma=0.697$, $\delta=0.414$ & \\
        Yakopcic MM \cite{updated_generalized_model} & $x_p=0.055$, $x_n=0.888$, $A_p=0.489$, $A_n=0.049$, $V_p=4.611$, $V_n=0.0$, $\gamma_1=0.714$, & 0.494\\
        & $\delta_1=0.409$, $\gamma_2=0.045$, $\delta_2=0.766$ & \\
        $q$-deformed MM & $x_p=0.21$, $x_n=0.571$, $A_p=0.321$, $A_n=0.049$, $V_p=4.543$, $V_n=0.0$, $\gamma_1=0.227$, & 0.457 \\
        & $\delta_1=1.021$, $\gamma_2=0.001$, $\delta_2=5.373$, $q=0.726$ & \\
        $q$-deformed MM state & $x_p=0.491$, $x_n=0.0$, $A_p=8.9$, $A_n=0.472$, $V_p=4.477$, $V_n=1.007$, $\gamma_1=0.002$, & 0.435\\
        & $\delta_1=20.623$, $\gamma_2=0.0$, $\delta_2=0.0$, $q=0.496$ & \\
        $q$-deformed M state & $x_p=0.492$, $x_n=0.25$, $A_p=9.105$, $A_n=0.262$, $V_p=4.487$, $V_n=0.0$, $\gamma=0.003$, & 0.431\\
        & $\delta=18.482$, $q=0.499$ & \\
        \hline\hline
    \end{tabular}
    \caption{All tested models with final parameters and NRMSEs}
    \label{table:fitting_results}
\end{table*}

\begin{lstfloat*}
	\caption{SPICE~\cite{spice_nagel} Implementation of $q$-deformed MM state model. Based on Ref.~\cite{generalized_memristive_spice_model}.}
	\label{lst:spice}
\begin{lstlisting}
* Q-Deformed Memristor SPICE model
* Base code provided by Yakopcic C. et.al.
* "Generalized Memristive Device SPICE Model and its
* Application in Circuit Design" (2013)

* Connections:
* TE  - top electrode
* BE  - bottom electrode
* XSV - External connection to plot state variable

.subckt Q-DeformedMM TE BE XSV

* Parameter vector
.params xp=0.491 xn=0.0 Ap=8.9 An=0.472 Vp=4.477 Vn=1.01
+gamma1=0.002 delta1=20.623 gamma2=0.0 delta2=0.0 q=0.496 xo=0.329

* Function EXPq(x) - Describes the q-deformed exp
.func EXPq(V1) {IF((q == 1),(exp(V1)),(IF(((1+(1-q)*V1) > 0),(pow((1+(1-q)*V1),(1/(1-q)))),(0))))}

* Function SINHq(x) - Describes the q-deformed sinh
.func SINHq(V1) {0.5*(EXPq(V1)-EXPq(-V1))}

* Multiplicitive functions to ensure zero state
* variable motion at memristor boundaries
.func wp(V) { (xp-V)/(1-xp)+1 }
.func wn(V) { V/(1-xn) }

* Function G(V(t)) - Describes the device threshold
.func G(V) {IF((V <= Vp),(IF((V >= -Vn),(0),(-An*(EXPq(-V)-EXPq(Vn))))),(Ap*(EXPq(V)-EXPq(Vp))))}

* Function F(V(t),x(t)) - Describes the SV motion 
.func F(V1,V2) {IF((V1 >= 0),(IF((V2 >= xp),(exp(-(V2-xp))*wp(V2)),(1))),
+(IF((V2 <= (1-xn)),(exp(V2+xn-1)*wn(V2)),(1))))}

* IV Response - Hyperbolic sine due to Ionic conduction
.func IVRel(V1,V2) {(V2*gamma1*SINHq(delta1*V1)+(1-V2)*gamma2*SINHq(delta2*V1))}

* Circuit to determine state variable
* dx/dt = F(V(t),x(t))*G(V(t))
Cx XSV 0 {1}
.ic V(XSV) = {xo}
Gx 0 XSV value={F(V(TE,BE),V(XSV,0))*G(V(TE,BE))}

* Current source for memristor IV response
Gm TE BE value = {IVRel(V(TE,BE),V(XSV,0))}

.ends Q-DeformedMM
\end{lstlisting}
\end{lstfloat*}

\clearpage
\bibliographystyle{elsarticle-num-names.bst} 
\bibliography{phys_a.bbl}

\end{document}